\documentclass[fleqn,usenatbib]{mnras}

\usepackage{newtxtext,newtxmath}

\usepackage[T1]{fontenc}

\DeclareRobustCommand{\VAN}[3]{#2}
\let\VANthebibliography\thebibliography
\def\thebibliography{\DeclareRobustCommand{\VAN}[3]{##3}\VANthebibliography}

\usepackage{graphicx}	
\usepackage{amsmath}	
\usepackage{xcolor}
\usepackage{hyperref}
\usepackage[normalem]{ulem}
\usepackage{caption}
\usepackage{subcaption}

\newcommand{\blip}{{\tt BLIP} }
\newcommand{\blips}{{\tt BLIP}'s\ }
\newcommand{\almax}{\ell_{\mathrm{max}}^a}
\newcommand{\blmax}{\ell_{\mathrm{max}}^b}
\newcommand{\nside}{{\tt nside} }

\title[SGWB in LISA from DWDs in the LMC]{A Stochastic Gravitational Wave Background in LISA from Unresolved White Dwarf Binaries in the Large Magellanic Cloud}

\author[S. Rieck et al.]{
\href{https://orcid.org/0009-0006-0978-7892}{Steven Rieck},$^{1,2}$\thanks{E-mail: riecksn@mail.uc.edu (SR); alexander.criswell@ligo.org (AWC)}\textsuperscript{\thanks{S. Rieck and A.W. Criswell contributed equally to this work and are co-first authors.}}
\href{https://orcid.org/0000-0002-9225-7756}{Alexander W. Criswell},$^{1 \star}$\textsuperscript{\footnotemark[2]}
\href{https://orcid.org/0000-0002-6725-5935}{Valeriya Korol},$^{3,4}$
\href{https://orcid.org/0000-0002-7743-2501}{Michael A. Keim},$^{5}$
\href{https://orcid.org/0000-0001-5489-4204}{Malachy Bloom},$^{6}$ \newauthor
\href{https://orcid.org/0000-0001-6333-8621}{Vuk Mandic}$^{1}$
\newauthor
\\
$^{1}$School of Physics and Astronomy, University of Minnesota, Minneapolis, MN 55455, USA\\
$^{2}$Department of Physics, University of Cincinnati, Cincinnati, OH 45221, USA\\
$^{3}$Max-Planck-Institut f{\"u}r Astrophysik, Karl-Schwarzschild-Straße 1, 85741 Garching, Germany\\
$^{4}$Institute for Gravitational Wave Astronomy \& School of Physics and Astronomy, University of Birmingham, Birmingham, B15 2TT, UK \\
$^{5}$Department of Astronomy, Yale University, PO Box 208101, New Haven, CT 06520-8101, USA\\
$^{6}$Department of Physics and Astronomy, Carleton College, Northfield, MN 55057, USA\\
}

\date{Accepted 2024 May 10. Received 2024 May 10; in original form 2023 September 2}

\pubyear{2024}

\begin{document}
\label{firstpage}
\pagerange{\pageref{firstpage}--\pageref{lastpage}}
\maketitle

\begin{abstract}
The Laser Interferometer Space Antenna (LISA) is expected to detect a wide variety of gravitational wave sources in the mHz band. Some of these signals will elude individual detection, instead contributing as confusion noise to one of several stochastic gravitational-wave backgrounds (SGWBs) -- notably including the `Galactic foreground', a loud signal resulting from the superposition of millions of unresolved double white dwarf binaries (DWDs) in the Milky Way. It is possible that similar, weaker SGWBs will be detectable from other DWD populations in the local Universe, including the Large Magellanic Cloud (LMC). We use the Bayesian LISA Inference Package ($\tt{BLIP}$) to investigate the possibility of an anisotropic SGWB generated by unresolved DWDs in the LMC. To do so, we compute the LMC SGWB from a realistic DWD population generated via binary population synthesis, simulate four years of time-domain data with \blip comprised of stochastic contributions from the LMC SGWB and the LISA detector noise, and analyze this data with \blips spherical harmonic anisotropic SGWB search. We also consider the case of spectral separation from the Galactic foreground. We present the results of these analyses and show, for the first time, that the unresolved DWDs in the LMC will comprise a significant SGWB for LISA.
\end{abstract}




\section{Introduction}
The launch of the Laser Interferometer Space Antenna (LISA) \citep{amaro-seoane_laser_2017} in 2035 will revolutionize gravitational wave (GW) astronomy. A space-based gravitational observatory, LISA will detect GWs in the millihertz frequency band, a range inaccessible to both pulsar timing arrays such as the International Pulsar Timing Array \citep[IPTA;][]{manchester_the_2013} and current ground-based detector networks such as the Laser Interferometer Gravitational-wave Observatory (LIGO), the Virgo detector, and the Kamioka Gravitational Wave Detector (KAGRA) \citep{aasi_advanced_2015,acernese_advanced_2015,adhikari_a_2020}.  

LISA is expected to detect a wide variety of astrophysical GW sources, including millions of double white dwarfs (DWDs) in the Milky Way (MW) and the nearby Universe \citep[for a review see][]{amaro-seoane_astrophysics_2022}. Some of these sources will be individually resolvable, whereas others will contribute to several potential stochastic gravitational wave backgrounds \citep[SGWBs; e.g.,][]{bonetti_gravitational_2020, babak_stochastic_2023,pozzoli_computation_2023c}. SGWBs arise from confusion noise formed by the overlap of many unresolved astrophysical or cosmological sources; evidence for such a signal in the nanohertz band has recently been detected \citep{agazie_the_2023a}. While many Galactic DWDs will be individually resolvable by LISA --- some serving as verification binaries for the instrument \citep[e.g.][]{Stroeer2006,Savalle2022,Finch2023} --- a far greater number of Galactic DWDs will contribute to a stochastic GW signal distributed mostly along the Galactic plane, comprised of the superposition of millions of individually unresolvable DWDs \citep{edlund_white_2005}. Characterization of this anisotropic MW foreground (so-called due to its prominence above the LISA detector noise) will be necessary in order to subtract it from the LISA data and identify other signals. Additionally, the foreground is of scientific interest in its own right as a means of studying MW structure and star formation history (SFH) \citep[e.g.][]{Benacquista2006,breivik_constraining_2020,georgousi_gravitational_2023}.

The MW is not the only host of DWDs detectable with LISA. Recent simulations show that nearby dwarf galaxies including the Large Magellanic Cloud (LMC), Small Magellanic Cloud, and Sagittarius Dwarf contain DWDs that will appear as individually resolvable LISA sources \citep{roebberf_milky_2020}. The number of resolvable DWDs depends on a dwarf galaxy’s mass, distance, and SFH \citep{korol_populations_2020}. However, as in the MW, the majority of DWDs in dwarf galaxies will not generate resolvable signals; for instance in the LMC population of ${\mathcal O}(10^6)$ DWDs only ${\mathcal O}(10^2)$ will be individually detectable \citep{korol_populations_2020}. It is possible that, as in the MW, these unresolved DWDs will contribute to an anisotropic SGWB detectable with LISA. To our knowledge, the detectability of SGWBs from nearby dwarf galaxies with LISA anisotropic SGWB searches has not been investigated prior to this work. 

Due to its high mass and relative proximity, the LMC is an ideal first candidate to evaluate the possibility of a SGWB from DWDs outside of the MW. Recent work has considered the detectability of individual DWDs in the LMC by constructing model populations based on hydrodynamic simulations and electromagnetic observations of its SFH and stellar density \citep{keim_large_2022}. Keim et al. found that while the LMC likely has only tens or hundreds of detectable DWDs with LISA signal-to-noise ratio (SNR) $>7$, it contains approximately two million DWDs in the LISA frequency band\footnote{The LMC is expected to contain approximately 61 million DWDs in total, but a vast majority are non-interacting with large orbital separations and are negligible sources of GWs. A frequency cut-off of $10^{-4} \; \mathrm{Hz}$ reduces this number to two million in the LISA frequency band (\citet{keim_large_2022}, by correspondence with the author).}. This quantity is significantly less than the DWDs in the MW; the LMC stellar mass \citep[e.g.][]{marel_new_2002} is roughly an order of magnitude less than the mass of the MW ($2.7\times10^9$\;M$_\odot$ vs several $10^{10}$\;M$_\odot$). At $\sim50 \; \mathrm{kpc}$ from Earth, the LMC signal is also reduced by distance, as GW amplitudes scale as the inverse of the distance. Given these considerations, we may expect the LMC SGWB to have approximately $1\%$ to $2\%$ the strength of the MW signal. On the other hand, while the MW DWDs are distributed across a large fraction of the sky, the LMC DWDs are focused in 77 square degrees, making the LMC a good target for an anisotropic SGWB search. 

In this work, we simulate and recover the SGWB signal in LISA from a model LMC population using the Bayesian LISA Inference Package (\texttt{BLIP}) \citep{banagiri_mapping_2021}. In \S\ref{sec:Methods} we describe the model population used to simulate the LMC signal and our code for simulation and recovery. Results are presented in \S\ref{sec:Results} and the conclusions of this study alongside possible future extensions are discussed in \S\ref{sec:discussion_conclusions}.

\section{Methods}
\label{sec:Methods}
To investigate the stochastic signal from the LMC we use $\tt{BLIP}$, described at length in \citet{banagiri_mapping_2021}. \blip is a Python package designed for the end-to-end simulation and Bayesian analysis of stochastic GW signals with LISA. In this study, we use the \blip spherical harmonic anisotropic stochastic search first presented in \citet{banagiri_mapping_2021}, which is explained in brief in \S\ref{sec:blip_sph}. \blip can simulate a wide variety of anisotropic stochastic GW signals; we make use of its capability to simulate a SGWB from a realistic simulated population of the unresolved DWDs in the LMC. This population is further described in \S\ref{sec:population}, while the simulated and recovered models in \blip are described in \S\ref{sec:BLIP_simulation} and \S\ref{sec:BLIP_recovery} respectively.

\subsection{Anisotropic SGWBs in \blip}\label{sec:blip_sph}
Simulation and recovery of anisotropic SGWBs in \blip is performed in the spherical harmonic basis. Several studies -- considering ground-based, space-based, and pulsar timing-based analyses -- have used versions of the spherical harmonic basis for expanding the sky distribution of GW power in e.g.~\citep{Cornish:2001hg, Ungarelli:2001xu, Kudoh:2004he, Taruya:2005yf, Taruya:2006kqa, Thrane:2009aa, Mingarelli:2013dsa, Taylor:2013esa, Renzini:2018vkx}. However, constraining the spherical harmonic distribution to be real and non-negative everywhere is a non-trivial problem that can hamper the accurate characterization of highly anisotropic sources such as the Galactic foreground --- or, indeed, the LMC. This is especially true for Bayesian analyses.

\citet{banagiri_mapping_2021} developed an explicitly Bayesian version of the spherical harmonic SGWB analysis for LISA wherein this problem was solved by fitting the square root of the GW power. Specifically, the spatial distribution of the SGWB on the sky (for purposes of both simulation and inference) is represented by the spherical harmonic coefficients $b_{\ell m}$. The $b_{\ell m}$s describe the spherical harmonic expansion of the \textit{square root} of the GW power on the sky $\mathcal{S}(\mathbf{n})$. The $b_{\ell m}$s are related to the usual spherical harmonic coefficients and functions of the GW power on the sky,\ $a_{\ell m}$ and $Y_{\ell m}$, respectively, via
\begin{equation}\label{eq:sph-harm}
\begin{split}
    \mathcal{S}(\mathbf{n}) = \sqrt{\mathcal{P}(\mathbf{n})} &= \left[\sum_{\ell=0}^{\ell_{\rm max}^a} \sum_{m = -\ell}^{\ell} a_{\ell m}Y_{\ell m}(\mathbf{n}) \right]^{1/2} \\
     &= \sum_{\ell=0}^{\ell_{\rm max}^b} \sum_{m = -\ell}^{\ell} b_{\ell m}Y_{\ell m}(\mathbf{n}),
\end{split}
\end{equation}
where $\ell_{\mathrm{max}}^b = \ell_{\mathrm{max}}^a /2$~\citep{banagiri_mapping_2021}. The $a_{\ell m}$ and $b_{\ell m}$ terms are directly related to each other via simple linear transformations involving Clebsch-Gordan coefficients \citep{banagiri_mapping_2021}. Characterizing the GW anisotropy in this way mathematically ensures that the GW power in every proposed sample is real and non-negative across the entire sky (see \citet{banagiri_mapping_2021} for details). 

In practice, the \blip anisotropic search infers and produces posterior distributions for each $b_{\ell m}$ coefficient (alongside spectral parameters; see \S\ref{sec:BLIP_recovery}) up to some $\ell_{\mathrm{max}}^b = \ell_{\mathrm{max}}^a /2$.\footnote{As the usual spherical harmonic $\ell_{\mathrm{max}}$ referred to in the literature is $\almax$, we will quote this truncation $\ell_{\mathrm{max}}$ in terms of $\almax$ throughout this work.} Using a higher $\almax$ for the anisotropic search increases the angular resolution of the search, but also increases the number of parameters that one must infer as $N_{\mathrm{par,sph}} \propto \almax (\almax + 1) / 2$. Additionally, as the \blip anisotropic search considers the LISA detector response to each spherical harmonic, the computational resources required to analyze data at large $\almax$ can become limiting. 

This latter point is also a limitation for simulation of anisotropic SGWBs with $\tt{BLIP}$, as the SGWB spatial distribution is simulated in the spherical harmonic basis. Accordingly, simulations of anisotropic signals in \blip similarly employ a truncation $\almax$. This, of course, results in highly-localized signals (like the LMC) spreading out over an area much larger than their true spatial extent on the sky. However, a study of BLIP's angular resolution (Bloom et al., in-prep) has shown that the value of $\almax$ used in the SGWB simulation does not impact the final spatial recovery so long as $\ell_{\mathrm{max,simulation}}^a \ge \ell_{\mathrm{max,recovery}}^a$. (Simply put, our analysis is insensitive to variations on smaller scales than it parameterizes, as one would expect intuitively.) Development work is ongoing to improve \blips performance for both simulation and analysis at high $\almax$ ($\gtrsim 8$), but these computational limitations remain relevant at present.

\subsection{Simulated LMC DWD Population}
\label{sec:population}
To date, no DWD has been observed in the LMC. Even within the MW most of the known LISA-detectable DWDs are found within a few kpc \citep[e.g.][]{kupfer_lisa_2024}; this is mainly due to the faint nature of white dwarf stars. Nonetheless, this highlights an opportunity for LISA to reveal the DWD population inaccessible to electromagnetic observatories as far as the LMC.
To model the LMC DWDs we employ a mock catalogue compiled by \citet{keim_large_2022}. It is based on a fiducial DWD population synthesis model computed with the {\sc SeBa} code \citep{SeBa,Toonen2012}, which has been calibrated on the observed DWDs (albeit in the Solar neighborhood) and, therefore, is in good agreement with the observed DWD space density and mass-ratio distribution \citep{Toonen2012, Toonen2017}.

Synthetic DWDs are distributed across the sky and assigned formation times and ages based on the Magellanic Clouds Photometric Survey and the observed, spatially resolved 2D SFH from \citet[][for a visual representation see their fig. 4]{2009AJ....138.1243H}. We refer to \citet[see their `Model 1']{keim_large_2022} for further details.

For the assumed LMC total stellar mass of $2.7 \times 10^9$\,M$_\odot$ \citep{2002AJ....124.2639V}, the adopted model yields $\sim 2\times10^6$ DWDs in the LISA frequency band. For this model, only about $\sim$500 DWD are individually resolved with SNR $>7$, assuming the mission lifetime of 4\,years with 100\% duty cycle. The detectable binaries have frequencies $>$1.7\,mHz (or equivalently binary orbital periods of $<$20 min) due to LISA's selection effects. The total number of LISA sources in the LMC  represents about $8\%$ of the MW DWD population. As detailed in \citet{keim_large_2022}, the difference between the two populations is twofold. Firstly, the number of LISA sources (and stars in general) scales linearly with the total mass of the host galaxy. The lower mass of the LMC thus decreases individual DWD detections. Secondly, unlike the MW, the LMC is an active site of star formation, and so a significant fraction of DWD in the adopted model have formed only $\sim\mathcal{O}(10^2)$\,Myr ago. This active star formation increases detections of individual DWDs in the LMC. 

\subsection{Simulated LISA Data}
\label{sec:BLIP_simulation}
The simulation of stochastic GW signals from DWD population-synthesis catalogues is a novel \blip feature demonstrated for the first time in this work. For each catalogue binary, we compute the (assumed monochromatic) strain amplitude from its binary masses and orbital frequency, following the conventions in \cite{wagg_legwork_2022}. We use the catalogue sky position and distance (as seen in the Solar System Barycentre frame) to bin the population in both frequency and sky direction. Binning on the sky is performed on a Healpix \citep{gorski_healpix_2005} map, with user-specified skymap pixel resolution, quantified by the Healpix $\tt{nside}$. In this work we use an \nside of 8 to generate our simulated signal. At our chosen skymap resolution, the area of each pixel equals approximately 53 square degrees. The angular size of the LMC is approximately 77 square degrees \citep{roebber_milky_2020}. Thus, in our initial simulated skymap the entire LMC is contained within only a few pixels. Simulating the LMC with a higher \nside would incur significantly higher computational cost for little-to-no ultimate effect due to limitations on the sky resolution of our analysis (see \S\ref{sec:blip_sph}). 

To compute the associated SGWB spectrum of the DWD population, we assume all DWD systems with individual SNR $>7$ are individually resolvable and can be subtracted from the data \citep{keim_large_2022}. We use \texttt{LEGWORK} \citep{wagg_legwork_2022} to calculate the SNR of every DWD considering the instrumental noise and Milky Way foreground given in \citet{robson_the_2019}, and remove from the population those DWDs with SNR $>7$. Disentangling the resolved and unresolved DWDs --- let alone the entire cacophony of LISA sources --- is beyond the scope of this work, requiring a global, simultaneous solution (e.g., \citealt{littenberg_prototype_2023}). We assume all other GW sources are perfectly characterized and subtracted from the data, and we first simulate a signal that includes only the unresolved LMC DWDs. In a second analysis, we also include a simple model of the MW foreground (see \S\ref{sec:MW_simulation}). Our simulation of the LMC DWDs is identical in each analysis. The monochromatic strains of the remaining unresolved binaries are then binned in frequency at a frequency resolution determined by the LISA nominal mission duration of 4 years, i.e. $\Delta f = 1/T_{\mathrm{obs}} \simeq 8\times10^{-9}~\mathrm{Hz}$. We consider a frequency range of $f \in [10^{-4},10^{-2}]$ Hz, as this will be the most-sensitive band of the LISA detector. 

\begin{figure*}
    \begin{subfigure}{1\columnwidth}
        \includegraphics[width=\columnwidth]{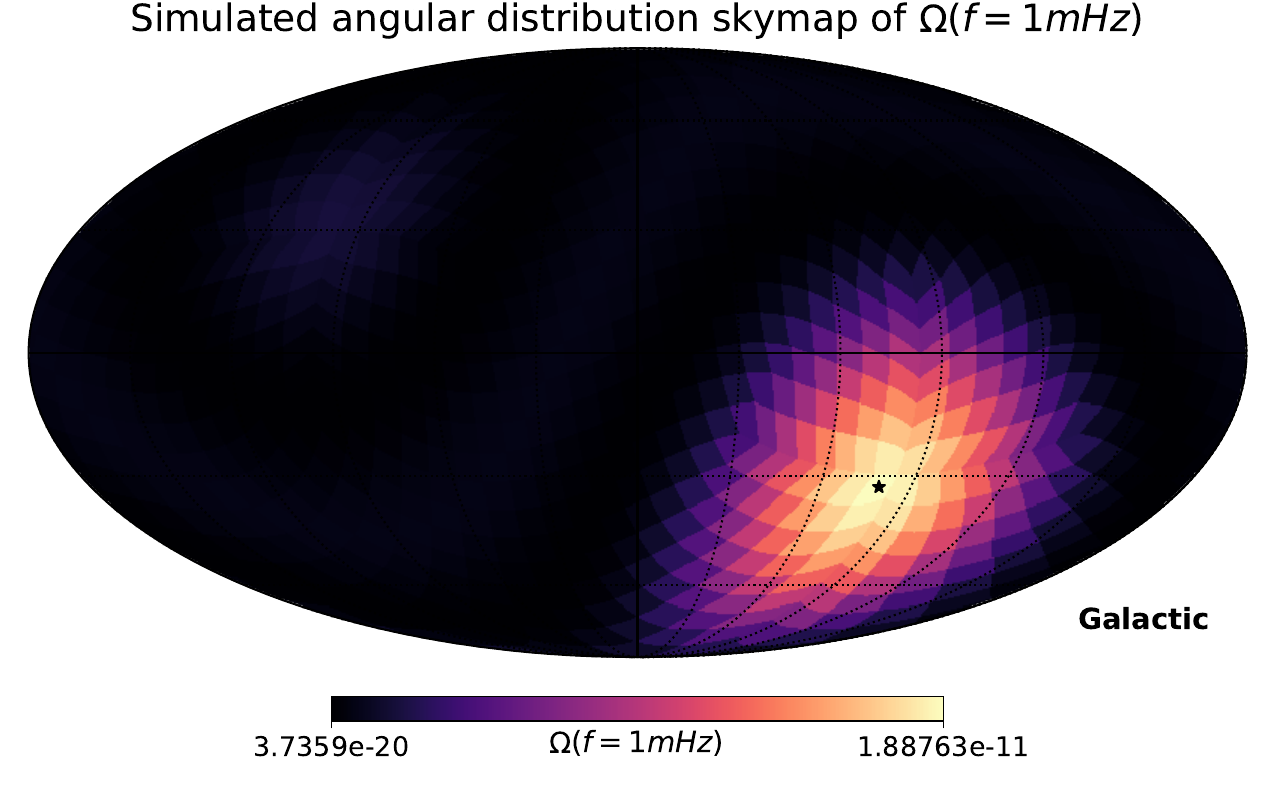}
        \subcaption[]{}
        \label{fig:simulation_skymap}
    \end{subfigure}
    \hfill
    \begin{subfigure}{1\columnwidth}
    	\includegraphics[width=\columnwidth]{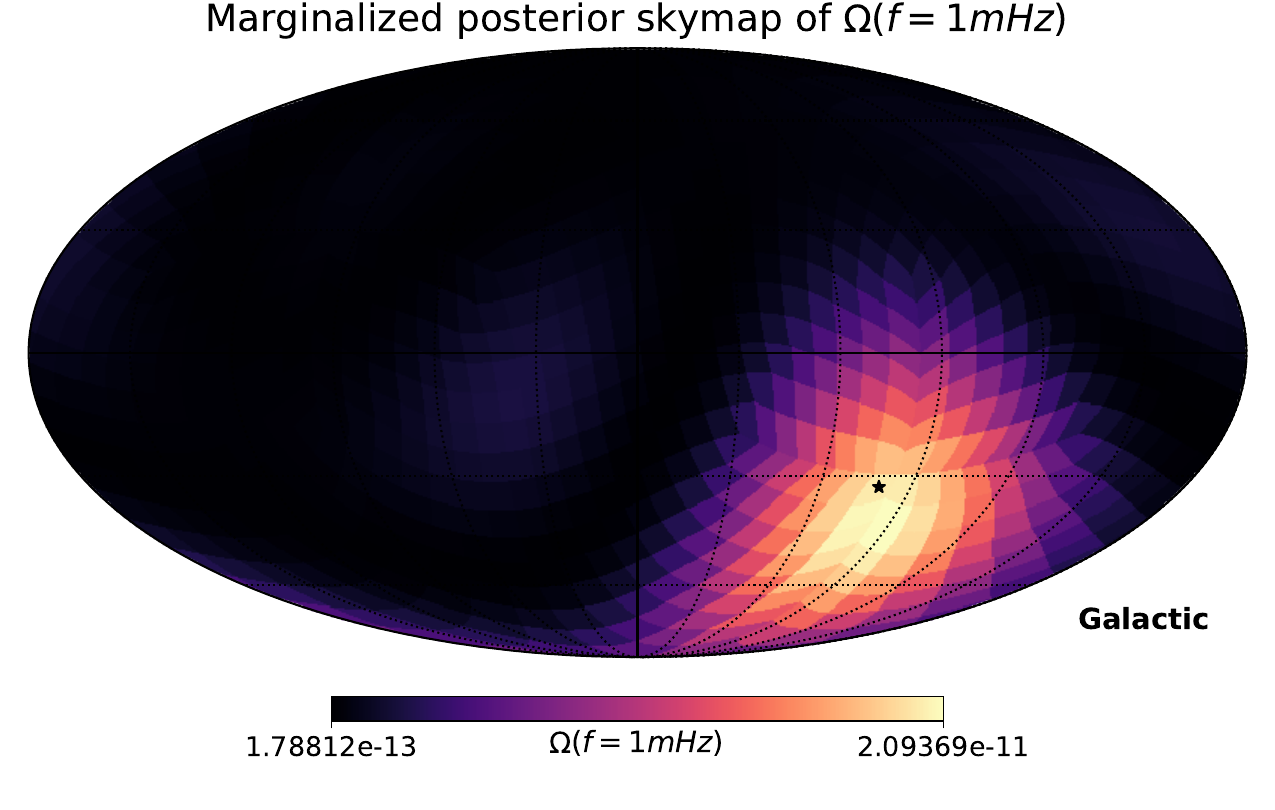}
        \subcaption[]{}
        \label{fig:posterior_skymap}
    \end{subfigure}
\caption{\textbf{(a)} The simulated sky distribution of $\Omega_{\mathrm{GW}}(1\mathrm{ mHz})$ for the LMC SGWB generated by our model DWD population. \textbf{(b)} The marginalized posterior sky distribution of $\Omega_{\mathrm{GW}}(1\mathrm{ mHz})$ inferred by our analysis of the LMC in isolation. Both skymaps are in the spherical harmonic basis at $\almax=4$ and display distribution of the dimensionless GW energy density $\Omega_{\mathrm{GW}}$ evaluated at 1 mHz. These skymaps do not include LISA instrumental noise. The black star marks the position of the LMC. The recovered sky distribution is consistent with both the simulated sky distribution and the position of the LMC.}
\end{figure*}

After the population skymap and spectrum are computed, \blip simulates a time series of the corresponding stochastic signal. It does so by computing the spherical harmonic representation of the population skymap up to some $\almax$ (we consider a simulation $\almax$ of 4 due to computational limitations; see \S\ref{sec:blip_sph}). \blip convolves both this spherical harmonic expansion and the population spectrum with the time-varying LISA response across frequency and all considered spherical harmonic modes (see \citealt{banagiri_mapping_2021} for details). Note that this process explicitly models the orbits of the LISA constellation and as such naturally accounts for the time-varying amplitude of highly anisotropic SGWBs like that of the LMC and the MW. The simulated population skymap as represented in the spherical harmonic basis can be found in Fig.~\ref{fig:simulation_skymap}.

The resulting GW time series is added to Gaussian detector noise with the spectral form given in the LISA proposal \citep{amaro-seoane_laser_2017}, reproduced below in Eqs.~\eqref{eq:posnoise} and~\eqref{eq:accnoise}, with $N_{p}=9\times10^{-42}$ and $N_{a}=3.6\times10^{-49}$\,Hz$^{-4}$ for the position and acceleration noise contributions, respectively:

\begin{equation}
    S_{p}(f) = N_{p} \left[ 1+\left( \frac{2mHz}{f}\right) ^{4} \right] Hz^{-1},
	\label{eq:posnoise}
\end{equation}
\begin{equation}
    S_{a}(f) = \left[1+\left( \frac{0.4 mHz}{f} \right)^{2}\right] \left[1+ \left( \frac{f}{8 mHz} \right) ^{4}\right] \times \frac{N_{a}}{ \left( 2\pi f \right) ^{4}}Hz^{-1}.
	\label{eq:accnoise}
\end{equation}

Throughout this study we simulate and model LISA data using the X --- Y --- Z time-delay interferometry (TDI) channels (see \citet{tinto_time-delay_2014} for a review of TDI in LISA). For further details on the \blip data simulation procedure, see \citet{banagiri_mapping_2021}. The simulated spectrum, as it appears in the detector, along with the simulated detector noise, is included in Fig.~\ref{fig:spectrum}. 

\label{fig:skymaps}

\subsubsection{Simple Milky Way Foreground}
\label{sec:MW_simulation}
We also include a simple analytic (i.e., non-population) simulation of the MW foreground. Its spatial distribution follows the simple bulge + disc model described in \citet{breivik_constraining_2020}. We use the "thin" model (see \citet{breivik_constraining_2020} for details), with radial scale height $r_{\mathrm{h}}=2.9$ kpc and vertical scale height $z_{\mathrm{h}}=0.3$ kpc. This simulated Galaxy is then used to create a skymap in the Solar System Barycentre frame as described in \S~6 of \citet{banagiri_mapping_2021}. As throughout the rest of this work, we represent this spatial distribution in the spherical harmonic basis. For the MW foreground spectrum, we use a tanh-truncated power law similar to that of (e.g.,) \citet{robson_the_2019}, such that
\begin{equation}
    \Omega(f) = \Omega_{\mathrm{ref}}\left(\frac{f}{f_{\mathrm{ref}}}\right)^{\alpha} \left(1+\tanh \left(\frac{f_{\mathrm{cut}} - f}{f_{\mathrm{scale}}}\right)\right),
	\label{eq:trunc_powerlaw}
\end{equation}
where for this simulation $\Omega_{\mathrm{ref}}=2\times10^{-5}$, $f_{\mathrm{ref}}=25$~Hz, $f_{\mathrm{cut}}=2$~mHz, and $f_{\mathrm{scale}}=0.4$~mHz. Although this more simplistic analytic function does not account for iterative subtraction of resolved MW DWDs, it remains a sufficient approximation for our purposes given the large uncertainties in the overall amplitude and shape of the MW foreground signal. This skymap and spectrum are then used to compute the GW time-series contribution of the MW foreground in the same manner as described above for the LMC.

\subsection{Model Recovery in \blip}
\label{sec:BLIP_recovery}
After generating the simulated data, \blip performs Bayesian parameter estimation via nested sampling with \texttt{dynesty} \citep{speagle_dynesty_2020}. This process is described in brief below; reference \citet{banagiri_mapping_2021} for a more detailed treatment. The \blip anisotropic search simultaneously models the LISA detector noise, the SGWB spectral distribution, and the SGWB spatial distribution, inferring posterior distributions for each of the parameters described below. 

LISA's instrumental noise is modelled in terms of the position and acceleration noise amplitudes $N_p$ and $N_a$, with the spectral form given by Eqs.~\eqref{eq:posnoise} and~\eqref{eq:accnoise}. We characterize the LMC SGWB spectrum using a power-law spectral model of the form
\begin{equation}
    \Omega(f) = \Omega_{\mathrm{ref}}\left(\frac{f}{f_{\mathrm{ref}}}\right)^{\alpha},
	\label{eq:powerlaw}
\end{equation}
where $\Omega_{\mathrm{ref}}=\Omega(f_{\mathrm{ref}}=25\ \mathrm{Hz})$ is the power-law amplitude at the reference frequency $f_{\rm ref}$ and $\alpha$ is the power-law spectral index (slope). The value of $f_{\rm ref}$ is an arbitrary choice. \blip recovers both $\Omega_{\mathrm{ref}}$ and $\alpha$ as free parameters. The majority of the LMC SGWB spectrum can be approximated as a power law, although this model will be unable to capture the high-frequency turnover in the spectrum; as this work focuses on establishing the LMC SGWB as a significant signal in LISA, more complex spectral models are left to future work (see \S\ref{sec:discussion_conclusions}). 

As discussed in \S\ref{sec:blip_sph}, the spatial distribution of the LMC SGWB on the sky is inferred in the spherical harmonic basis. Our final spatial posteriors are given in terms of the $b_{\ell m}$s, from which it is straightforward to compute the corresponding $a_{\ell m}$s and SGWB power skymap. We choose an analysis $\almax$ of 4, in keeping with our choice for the simulated LMC spatial distribution. 

The Fourier-domain likelihood used in \blips nested sampling is a complex multivariate Gaussian \citep{adams_discriminating_2010} whose covariance is a function of the parameters in the previous four equations: $\mathcal{L}(\tilde{d}|N_{p},N_{a},\Omega_{\mathrm{ref}},\alpha,\{b_{\ell,m}\})$. The likelihood is given by Eq.~$32$ from \cite{banagiri_mapping_2021}:

\begin{equation}
    \label{eq:likelihood}
    \begin{split}
    &\mathcal{L}(\tilde{d}|N_{p},N_{a},\Omega_{\mathrm{ref}},\alpha,\{b_{\ell,m}\}) = \\ &\prod_{f,t} \frac{1}{2\pi T_{\mathrm{seg}}|C_{IJ}(f,t)|}\times \mathrm{exp} \left( - \frac{2 \tilde{d}^{*}_{f,t} C_{IJ}(f,t)^{-1} \tilde{d}_{f,t}}{T_{\mathrm{seg}}}    \right)
    \end{split}
\end{equation}
where $T_{\mathrm{seg}}$ is the length of each time segment, $C_{IJ}(f,t)$ is the channel covariance matrix, and $\tilde{d}_{f,t}$ is the array of data in the Fourier domain for the three LISA channels measured in the time segment labelled by $t$ and at frequency $f$. For explicit definitions of these terms see discussion in \cite{banagiri_mapping_2021} and original derivations in \citet{cornish_detecting_2001,cornish_space_2001}.

\subsubsection{Joint Model with the Milky Way Foreground}\label{sec:LMC+MW_methods}
We also consider a joint model that simultaneously infers the LMC SGWB alongside the MW foreground. This is a simplified, prototype demonstration of the full, flexible spectral separation infrastructure developed for \blip (Criswell et al., in prep.). Accordingly, we restrict ourselves to a simple MW model: we assume the MW spatial distribution is well-measured \textit{a priori} from the resolved Galactic DWDs, and fix its skymap to the analytic distribution described in \S\ref{sec:MW_simulation}. We then use the spectral model of Eq.~\eqref{eq:trunc_powerlaw}, fixing $f_{\mathrm{scale}}=0.4$ mHz, and inferring the set of free parameters $\vec{\theta}_{\mathrm{MW}} = \{\Omega_{\mathrm{ref,MW}},\alpha_{\mathrm{MW}},f_{\mathrm{cut}}\}$. The joint likelihood is then 
$\mathcal{L}(\tilde{d}|\vec{\theta}_{\mathrm{n}}; \vec{\theta}_{\mathrm{LMC}};\vec{\theta}_{\mathrm{MW}})$, where $\vec{\theta}_{\mathrm{n}}=\{N_{p},N_{a}\}$ describe the noise and $\vec{\theta}_{\mathrm{LMC}}=\{\Omega_{\mathrm{ref}},\alpha,\{b_{\ell,m}\}\}$ describe the LMC as discussed above. We leave a full discussion of \blips approach to spectral separation to Criswell et al. (in prep.). We stress that this simple model is a first pass at resolving the LMC SGWB in the presence of the MW foreground. A detailed treatment of spectral separation between the LMC and MW signals is sufficiently involved so as to warrant its own dedicated study.\footnote{See \S\ref{sec:discussion_conclusions} for further discussion as to what such a study could entail.} As such, more complicated models are outside the scope of this initial work, which primarily seeks to establish the LMC SGWB as a significant stochastic contribution in LISA.

\section{Results}
\label{sec:Results}
We include results from two simulations. In the first, we simulate the LMC SGWB generated from the population described in \S\ref{sec:population} with LISA instrumental noise. In \S\ref{sec:LMC_alone_recovery} we present the results of the recovery process described in \S\ref{sec:BLIP_recovery}. In \S\ref{sec:LMC+MW_recovery} we present a recovery of the LMC in the presence of a simple realization of the MW foreground, as described in \S\ref{sec:LMC+MW_methods}. 

\subsection{LMC SGWB Spectrum}
\begin{figure}
	\includegraphics[width=\columnwidth]{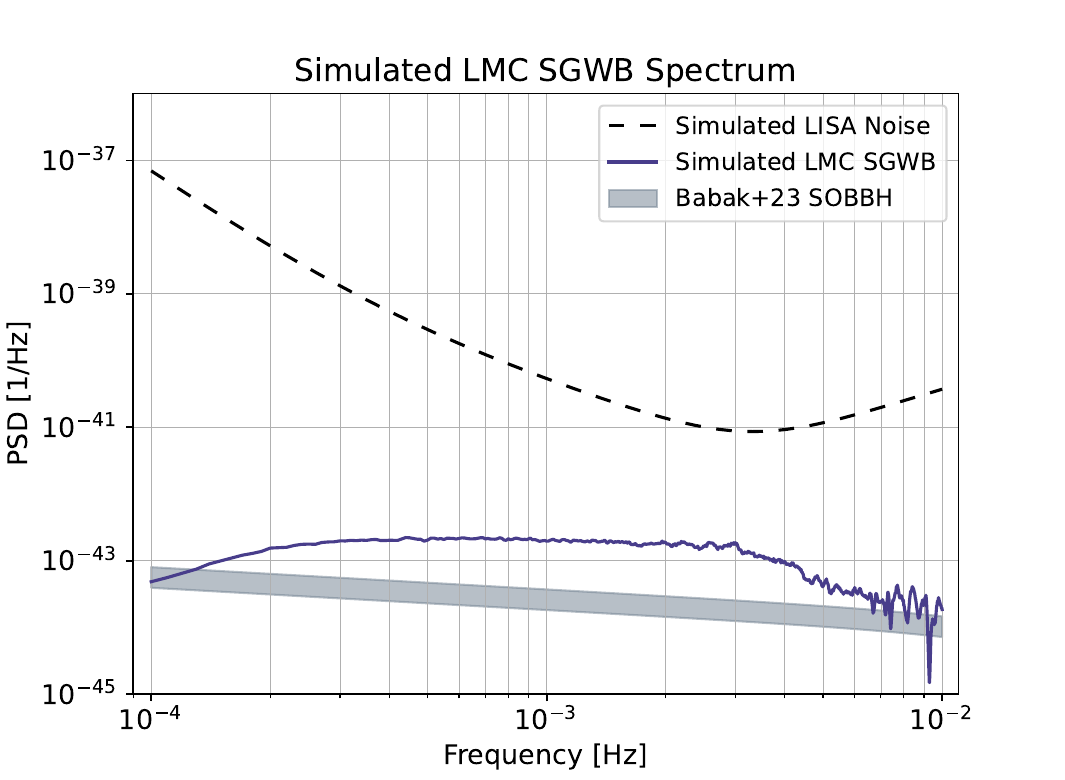}
    \caption{The simulated, population-derived LMC SGWB PSD. The LISA instrumental noise spectrum and the \citet{babak_stochastic_2023} interquartile prediction for the LISA SOBBH SGWB are shown for reference. Both SGWB PSDs are shown convolved with the LISA response. Note that the LMC SGWB amplitude exceeds that of the SOBBH signal in the relevant frequency band.}
    \label{fig:lmc_sobbh}
\end{figure}

The population-derived power spectrum of the LMC SGWB is shown in Fig.~\ref{fig:lmc_sobbh}. Notably, the amplitude of the LMC signal is comparable to --- and even exceeds --- that of the expected SGWB from extragalactic stellar-origin binary black holes (SOBBHs), shown here using the observationally-driven estimate of \citet{babak_stochastic_2023}. The LMC signal will therefore comprise a significant SGWB for LISA, and will be important to consider in efforts to characterise the SOBBH SGWB and other underlying SGWBs. This result is the first demonstration of the LMC SGWB as a relevant signal for LISA.

\subsection{Recovery of the LMC SGWB in Isolation}\label{sec:LMC_alone_recovery}
\begin{figure}
	\includegraphics[width=\columnwidth]{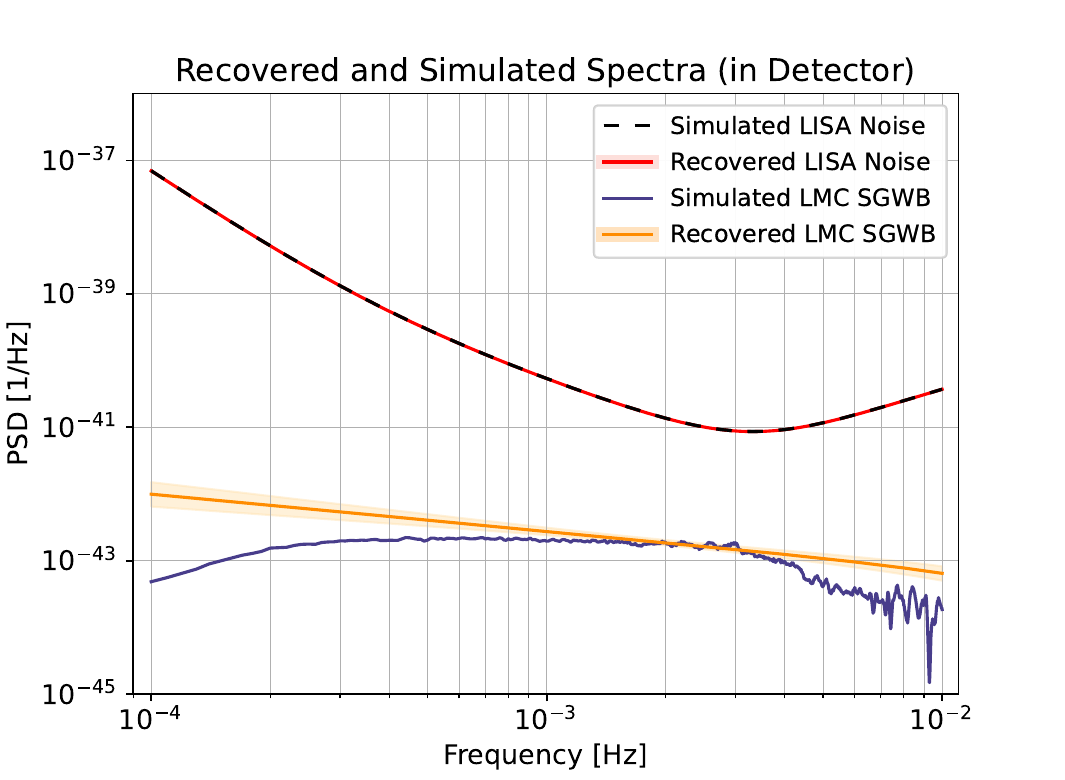}
    \caption{The simulated and inferred power spectral density of the LMC SGWB and the LISA detector noise. For the inferred spectra, the solid lines and shaded regions are the median and 95\% credible intervals, respectively, of the marginalized posterior spectral fit. As can be seen in Fig.~\ref{fig:corners_lmc_alone}, the noise spectrum is recovered extremely precisely; as a result the 95\% credible intervals are difficult to see by eye. Note that the power-law spectral fit has highest fidelity to the simulated LMC spectrum over the sensitive band of 1-4 mHz. }
    \label{fig:spectrum}
\end{figure}

We present here an analysis of the LMC SGWB in isolation (i.e., assuming the MW foreground has been subtracted) using an integration time of $1.26\times10^{8}$ seconds, approximately the planned LISA mission duration of four years, and considering a frequency band of $f \in [10^{-4},10^{-2}]$ Hz. We simulate and recover the LMC SGWB in the spherical harmonic basis, use a power law to model the SGWB spectrum, and model the LISA detector noise according to the spectral form given in Eqs.~\eqref{eq:posnoise} and~\eqref{eq:accnoise}. The corresponding marginalized posterior skymap computed from the inferred $b_{\ell m}$s is shown in Fig.~\ref{fig:posterior_skymap}, and the marginalized posterior detector-convolved power spectral density (PSD) is shown in Fig.~\ref{fig:spectrum} (alongside the PSDs of the simulated detector noise and of the SGWB due to the LMC DWD population). Posterior samples for all parameters are shown in Fig.~\ref{fig:corners_lmc_alone}.

As seen in Fig.~\ref{fig:posterior_skymap}, the inferred distribution of power on the sky is consistent with both the true position of the LMC and the simulated LMC SGWB skymap (Fig.~\ref{fig:simulation_skymap}). While more precise localization of the LMC SGWB could in principle be achieved with higher $\almax$ or a targeted directional search that takes advantage of the known position of the LMC, we leave these avenues of exploration to future work.

The inferred power-law spectrum of the LMC SGWB is shown in Fig.~\ref{fig:spectrum}, alongside the simulated and inferred noise spectra and the simulated population-derived spectrum of the LMC SGWB. The inferred amplitude and slope of the power-law model used in this study are most impacted by the shape of the LMC spectrum at frequencies where its SNR is largest --- namely 1-4 mHz, where the simulated LMC spectrum is closest to the LISA noise curve. At frequencies outside this range, the power-law model does not adequately describe the complexity of the simulated LMC SGWB spectrum, and hence it overestimates the contribution from the LMC signal at these frequencies. We leave treatment of more complex or nonparametric spectral models to future work, although we note that the overall low SNR of the LMC may make constraining highly-complex models difficult (unless the dimensionality of the inference problem is otherwise reduced by, for example, a targeted directional search). The noise spectra is recovered extremely well, due to the fact that we recover it using the exact functional form that we initially simulate. Ultimately the noise spectral shape will not be precisely known, which will introduce additional error. 

Finally, we perform model comparison via Bayes factor and consider two cases: our power-law spherical harmonic model including LISA noise and the LMC SGWB, and a noise-only model. Using the same four-year dataset including the LMC SGWB described in \S\ref{sec:BLIP_simulation}, we repeat our analysis using a model that only accounts for the LISA detector noise in terms of $N_p$ and $N_a$ as given in Eqs.~\eqref{eq:posnoise} and~\eqref{eq:accnoise} (neglecting the presence of any kind of underlying SGWB). Computing the Bayesian evidences of each model ($\mathcal{Z}_1$ for the noise + SGWB model; $\mathcal{Z}_2$ for the noise-only model) is trivial due to our use of nested sampling via {\tt dynesty}, which produces the Bayesian evidence as its primary product \citep{speagle_dynesty_2020}. We compute the log Bayes factor to be
$$\log K = \log\mathcal{Z}_1 - \log\mathcal{Z}_2 = 310 \pm 3,$$
constituting decisive evidence\footnote{For reference, a log Bayes factor of $1$ is substantial to strong evidence, and any log Bayes factor $>2$ is typically considered decisive evidence in favor of one model over another \citep{kass_bayes_1995}.} in favor of our SGWB plus noise model over the noise-only model. We conclude that --- in the absence of the MW foreground signal and for the case of stationary, Gaussian noise with a fixed, equilateral LISA constellation --- we are able to detect and characterize the LMC SGWB signal. Relaxing any of these assumptions will reduce LISA's sensitivity to SGWBs \citep[see, e.g.,][]{hartwig_stochastic_2023,muratore_impact_2024} and, accordingly, impact the ability of the LISA to detect and characterize the LMC SGWB. While fully accounting for these factors is beyond the scope of this work, we present a simplified treatment of a search for the LMC SGWB in the presence of the MW foreground in the following section.

\subsection{Recovery of the LMC SGWB with the MW Foreground}\label{sec:LMC+MW_recovery}

We now turn to the case of the LMC SGWB in the presence of the MW foreground. We additionally include in our simulated data a simple MW foreground as described in \S\ref{sec:MW_simulation}; the simulation procedure for the LISA instrumental noise and LMC SGWB is otherwise unchanged. This new dataset is then analyzed with the joint inference model described in \S\ref{sec:LMC+MW_methods}; all other quantities of interest (integration time, frequency range, etc.) are identical to the procedure described in \S\ref{sec:LMC_alone_recovery} for the LMC in isolation.

\begin{figure}
	\includegraphics[width=\columnwidth]{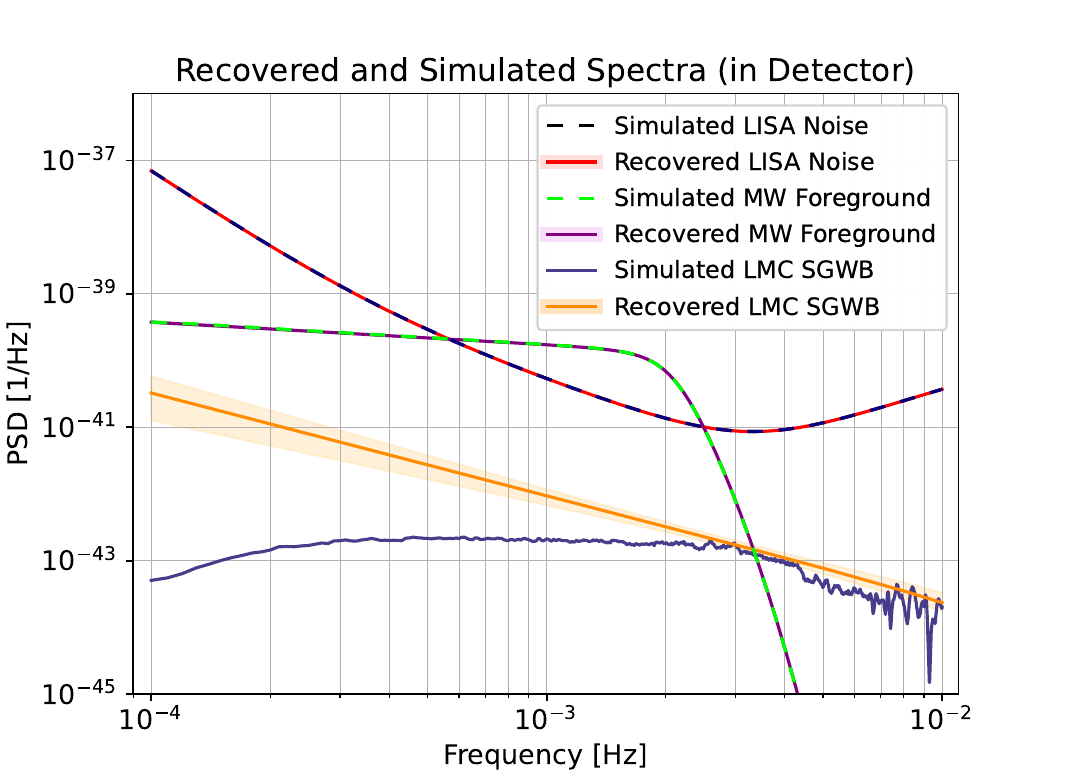}
    \caption{The simulated and inferred PSDs of the LMC SGWB, the MW foreground, and the LISA detector noise. For the inferred spectra, the solid lines and shaded regions are the median and 95\% credible intervals, respectively, of the marginalized posterior spectral fit. The precise recovery of both the LISA noise and MW foreground renders their respective medians and 95\% credible intervals nearly indistinguishable. The simple power law model for the LMC signal again results in an overestimation of power at low frequencies. The signal is most accurately recovered above 3 mHz where the contribution from the MW foreground is minimal.}
    \label{fig:MW+LMC_spectrum}
\end{figure}

We find that, despite the presence of the MW foreground, we are again able to detect and characterize the simulated LMC SGWB. The recovered spectral distribution of the LMC SGWB in the presence of the MW foreground is shown in Fig.~\ref{fig:MW+LMC_spectrum}, alongside those of the noise and the MW foreground. As before, we display the simulated and inferred spectra for each of our model components. Our recovered model successfully describes the LISA instrumental noise, MW foreground, and LMC SGWB simultaneously. Posterior samples for all spectral parameters are shown in Fig.~\ref{fig:corners_lmc_mw_spectral}. The presence of the MW does affect the recovered LMC SGWB, reducing the recovery quality below $\sim3$ mHz causing the power law to even more dramatically overestimate the LMC SGWB. Above $\sim3$ mHz, the recovered power law follows closely above the simulated LMC SGWB spectrum. It is again clear that the majority of information is being gleaned from the region around $\sim3$ mHz where the LMC SNR would be highest; more refined spectral models may be able to leverage this fact in future.

\begin{figure}
	\includegraphics[width=\columnwidth]{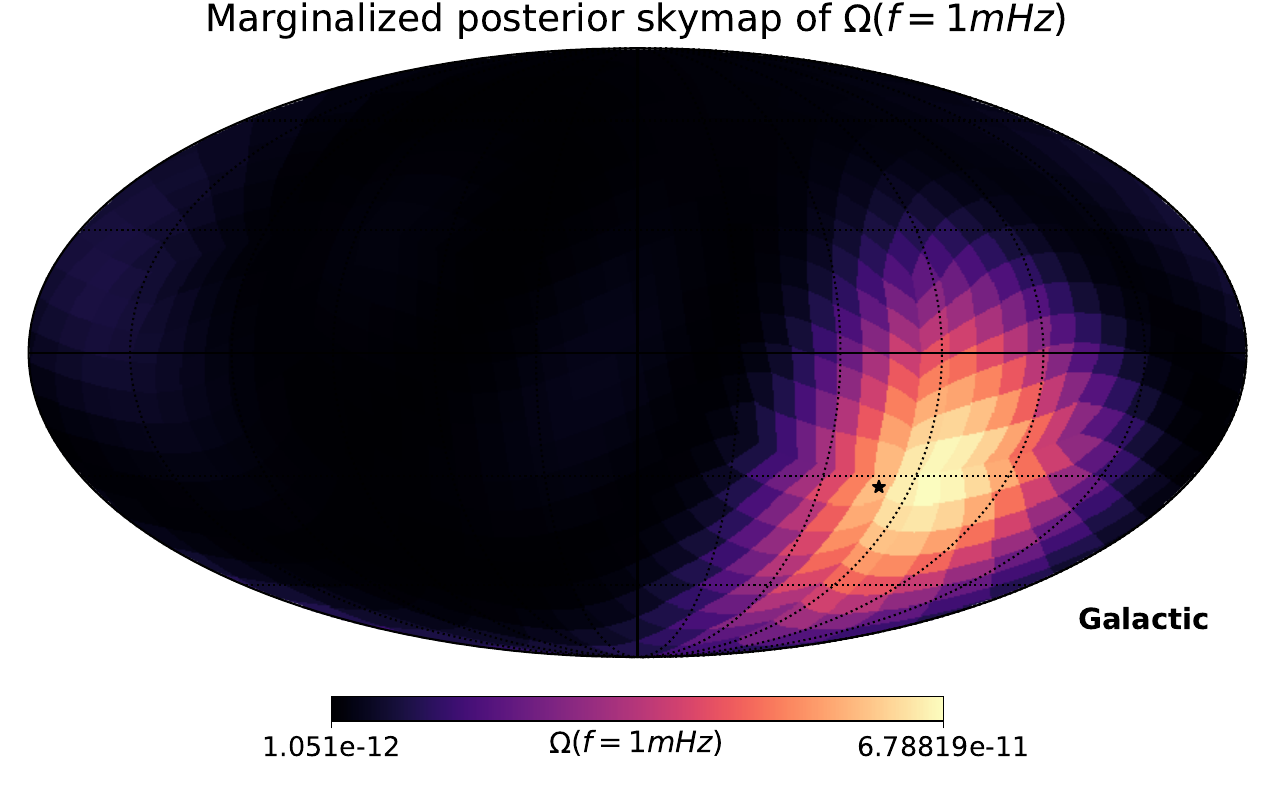}
    \caption{The marginalized posterior sky distribution of $\Omega_{\mathrm{GW}}(1\mathrm{ mHz})$ inferred by our analysis for the LMC SGWB in the presence of the MW foreground. The simulated LMC in this simulation is identical to Fig.~\ref{fig:simulation_skymap}. We represent the signal in the spherical harmonic basis at $\almax=4$. This skymap does not include LISA instrumental noise or the MW foreground, though both are present in the simulation. The black star marks the true position of the LMC. The recovered sky distribution is consistent with both the simulated signal and the true position of the LMC.}
    \label{fig:MW+LMC_skymap}
\end{figure} 

The LMC SGWB spatial recovery in the presence of the MW foreground can be seen in Fig.~\ref{fig:MW+LMC_skymap}.  It is important to note that this figure only displays the inferred distribution of power on the sky (i.e. the spherical harmonic spatial model for the LMC SGWB), and does not include the contribution from the MW (which is assumed known and therefore not inferred; see \S\ref{sec:LMC+MW_methods}). The associated posterior samples are shown in Fig.~\ref{fig:corners_lmc_mw_spatial}. As would be expected, the quality of the spatial recovery is degraded somewhat in the presence of the MW (and with a more statistically-complex signal model). While the extent of the inferred LMC spatial distribution is similar to the simulated skymap and the true position of the LMC is included in our recovered spatial distribution, it does experience some bias, shifting slightly off of the true position of the LMC.

Finally, we again perform a second analysis of the same simulated LISA data (including LISA instrumental noise, the MW foreground, and the LMC SGWB) with a model which accounts for LISA instrumental noise and the MW foreground, but neglects the presence of the LMC. We compute the log Bayes factor $(\mathrm{log}K)$ for this case using the Bayesian evidences of each model ($\mathcal{Z}_1$ for the LMC-included model; $\mathcal{Z}_2$ for the LMC-absent model):
$$\log K = \log\mathcal{Z}_1 - \log\mathcal{Z}_2 = 92 \pm 4,$$
While this Bayes factor is reduced compared to that for the LMC in isolation --- as expected, the MW foreground makes the LMC SGWB more difficult to recover --- it still constitutes extremely decisive evidence in favor of the model that includes the LMC SGWB.

\section{Discussion and Conclusions}
\label{sec:discussion_conclusions}
In this work, we evaluate for the first time the existence and prospects for LISA of an anisotropic SGWB arising from the unresolved DWDs in the LMC. We use a population catalog generated using realistic stellar synthesis codes to create a model of the LMC, which we then use to simulate its DWD-generated SGWB with $\tt{BLIP}$. We use \blips spherical harmonic, Bayesian search for anisotropic SGWBs to demonstrate a proof-of-concept recovery of the LMC SGWB both in isolation and in the presence of the MW foreground.

We find that the simulated SGWB from the unresolved DWDs in the LMC can be recovered in the presence of LISA instrumental noise using \blip with four years of integration time and a power-law spherical harmonic signal model. Model comparison between the noise + SGWB power-law spherical harmonic model and a noise-only model yields decisive evidence in favor of the presence of the LMC SGWB signal. The recovered position of the LMC on the sky is consistent with its true location, and the LMC SGWB spectrum can be well modeled as a simple power law over the sensitive frequency band (roughly 1-4 mHz). 

Additionally, we find that we are able to simultaneously recover the LMC SGWB and a rudimentary model of the MW foreground. While the presence of the MW has a noticeable, adverse effect on the recovery of the LMC SGWB, the recovered spatial distribution remains consistent with the true position of the LMC, and our power-law spectral model only slightly overestimates the LMC spectrum above $~3$ mHz. As in the LMC-only case, model comparison via Bayes factor yields decisive evidence in favor of the presence of the LMC SGWB signal. While a detailed treatment of spectral separation between realistic, population-derived realizations of the MW and LMC signals is required to make a strong statement of detectability --- and remains a subject of future work --- this result is nonetheless extremely promising for the prospects of LISA to detect and characterize the LMC SGWB.

While the power-law spectral model employed here is accurate to the simulated LMC spectrum where the LISA noise curve is lowest and the MW foreground has dropped off, outside these areas, it does not capture the full spectral shape of the LMC SGWB. Further characterization of the LMC SGWB with more complex spectral and/or spatial models is one promising avenue of future work. One could, for example, leverage the known location of the LMC to infer only its spectral distribution while holding its spatial distribution fixed, thereby reducing model complexity along one axis and allowing for (e.g.) a truncated or broken power-law spectral model to better capture the cutoff in the LMC SGWB spectrum. Such a model could also be informed by our theoretical knowledge of the LMC SGWB, either by setting astrophysically-motivated priors on its parameters, or fixing those parameters that see little variation across different population-synthesis realizations of the LMC. Conversely, ongoing efforts to incorporate non-parametric spectral models into \blip could enable more accurate characterization of the LMC spectrum, at the cost of increased difficulty of spectral separation from the MW foreground. With more precise spectral models, it may be possible to characterize the LMC SGWB well enough to gain information about the distribution of DWDs in the LMC and learn about its structure, mass, and/or SFH. Methods have been proposed to study the MW in this way using the unresolved Galactic DWDs \citep[e.g.][]{breivik_constraining_2020}, so it is possible that similar techniques could be used to study the LMC. In particular, it may be possible to achieve a measurement of the LMC mass via a similar approach to the one described in \cite{korol_mass}, which used the resolvable binaries in the LMC. Additionally, the analysis presented in this work is generalizable to simulation and recovery of the (albeit weaker) SGWBs from the Small Magellanic Cloud and other dwarf galaxy satellites of the Milky Way.

Finally, the development of refined approaches to concurrent characterization of the LMC SGWB and the MW foreground will be vital moving forward. The spectral overlap between these signals is significant; neglecting to properly account for the LMC SGWB could lead to spectral biases for analyses of the MW foreground. Despite their close proximity in terms of LISA's angular resolution, the spatial distributions of the MW and LMC are distinct on the sky and --- as demonstrated in this work --- can be used to aid in spectral separation between these signals. In particular, the spatial distribution of the LMC on the sky is well known from electromagnetic observations; our anisotropic search at high $\almax$ and/or a targeted directional search could leverage this fact. One could also incorporate concurrent GW localization measurements of the resolved DWDs in the LMC, improving prospects for resolving the LMC SGWB by jointly modelling the 3D spatial distribution of the LMC population (as has been proposed for the MW population \citep{adams_astrophysical_2012}). Finally, a pixel-basis method to describe the spatial distribution of a signal provides a promising alternative to a spherical-harmonic basis approach, which by necessity describes the entire sky rather than the region containing the LMC specifically. This method would be well-suited to enabling realistic spectral separation of the stochastic contributions from unresolved MW and LMC DWDs. 

Proper, joint treatment of both the LMC SGWB and MW foreground will likely be crucial for detecting and characterizing other, lower-amplitude SGWBs. The SOBBH background \citep[e.g.,][]{babak_stochastic_2023} is likely of comparable or lower amplitude in comparison to the LMC SGWB; see Fig.~\ref{fig:lmc_sobbh}. Characterization of the LMC SGWB is thus extremely relevant when considering the search for the SOBBH SGWB, as well as other, underlying backgrounds --- including those of cosmological origin.

\section*{Acknowledgements}
This work is supported by the NASA grant 90NSSC19K0318, and utilized computing resources provided by the Minnesota Supercomputing Institute at the University of Minnesota. Packages used for this work include {\tt numpy} \citep{harris_array_2020}, {\tt scipy} \citep{virtanen_scipy_2020a}, {\tt ChainConsumer} \citep{ChainConsumer}, and {\tt Matplotlib} \citep{hunter_matplotlib_2007}. The authors would like to thank Sharan Banagiri, Joe Romano, and Jessica Lawrence for their work on BLIP and many helpful conversations, as well as the anonymous reviewer for their thorough and insightful comments and suggestions.

\section*{Data Availability}

All data and code used in this study is publicly available. The simulated LMC population data along with all generated SGWB data and resulting posterior distributions are available at https://zenodo.org/records/10783952. The \blip package is open source and is available at https://github.com/sharanbngr/blip.


\bibliographystyle{mnras}
\bibliography{bib_final} 

\newpage
\onecolumn
\appendix
\section{Additional Figures}
\begin{figure*}
	\includegraphics[width=\textwidth]{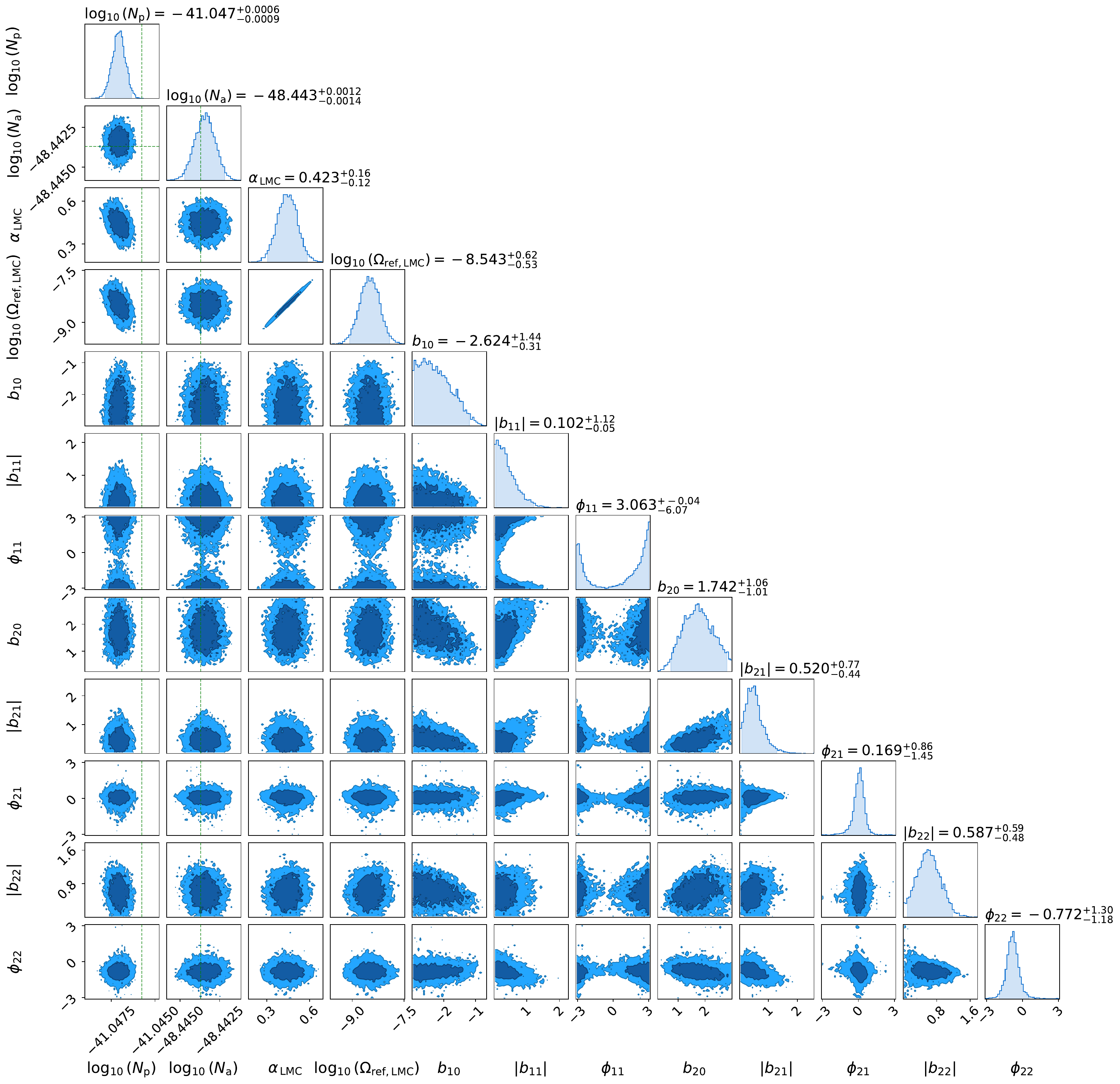}
    \caption{Corner plot for the analysis in \S\ref{sec:LMC_alone_recovery} of the LMC SGWB in isolation, showing the 1- and 2-dimensional marginalized posterior samples for each of our model parameters. These are (moving left to right): the LISA position and acceleration noise amplitudes ($\mathrm{log}_{10}(N_{\mathrm{p}})$ and $\mathrm{log}_{10}(N_{\mathrm{p}})$, respectively); the SGWB power-law model slope ($\alpha$) and log amplitude ($\log_{10}(\Omega_{\mathrm{ref}})$); and the magnitude and phase of the $b_{\ell m}$ spherical harmonic coefficients up to $\blmax=2$ ($\almax=4$). The true values of the noise parameters are marked with green dashed lines. The remaining parameters do not have defined true values, as our simulated signal is generated from a DWD population. Contours shown are 1- and $2\sigma$. A careful eye will note a slight bias in the recovery of the position noise contribution, $N_{\mathrm{p}}$. This is a result of our power-law spectral model being an imperfect fit for the population-derived, non-power-law spectrum of the LMC SGWB; repeating this study without the inclusion of the LMC signal results in unbiased noise recoveries. Potential future approaches to fitting the LMC signal with higher fidelity are discussed in \S\ref{sec:discussion_conclusions}.}
    \label{fig:corners_lmc_alone}
\end{figure*}

\begin{figure*}
	\includegraphics[width=\textwidth]{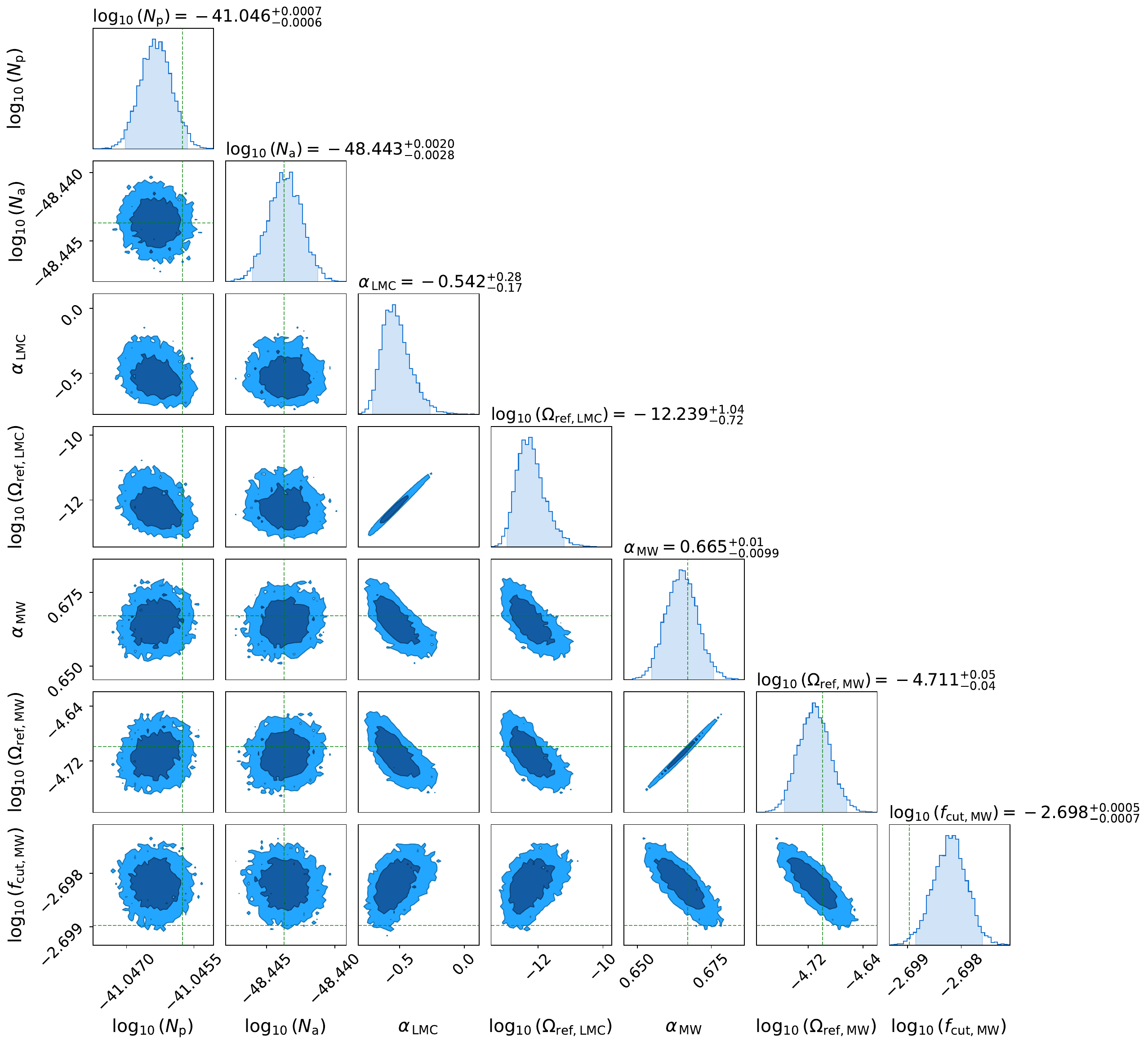}
    \caption{Spectral parameters corner plot for the analysis in \S\ref{sec:LMC+MW_recovery} of the LMC SGWB alongside a simple simulation of the MW foreground, showing the 1- and 2-dimensional marginalized posterior samples for all spectral model parameters. Spatial parameter samples are shown in Fig.~\ref{fig:corners_lmc_mw_spatial}. Included parameters are (moving left to right): the LISA position and acceleration noise amplitudes ($\mathrm{log}_{10}(N_{p})$ and $\mathrm{log}_{10}(N_{a})$, respectively); the LMC SGWB power-law model slope ($\alpha_{\mathrm{LMC}}$) and log amplitude ($\log_{10}(\Omega_{\mathrm{ref,LMC}})$); and the MW foreground truncated power-law model slope ($\alpha_{\mathrm{MW}}$), log amplitude ($\log_{10}(\Omega_{\mathrm{ref,MW}})$), and log cutoff frequency ($\log_{10}(f_{\mathrm{cut,MW}})$). True values are marked with green dashed lines. As before, the LMC model parameters do not have defined true values. Contours shown are 1- and $2\sigma$.}
    \label{fig:corners_lmc_mw_spectral}
\end{figure*}

\begin{figure*}
	\includegraphics[width=\textwidth]{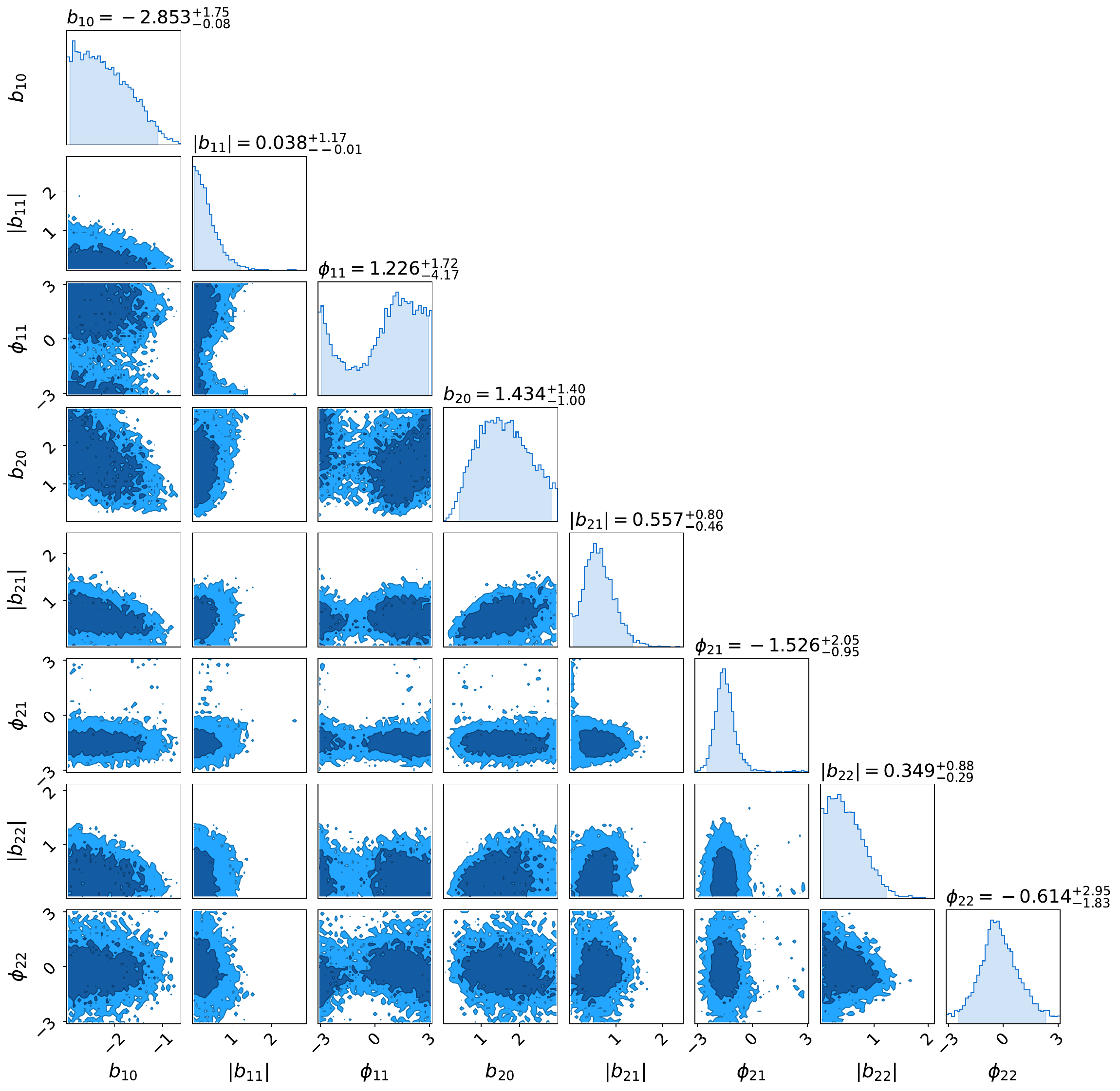}
    \caption{Spatial parameters corner plot for the analysis in \S\ref{sec:LMC+MW_recovery} of the LMC SGWB alongside a simple simulation of the MW foreground, showing the 1- and 2-dimensional marginalized posterior samples for the LMC spatial model parameters (the MW spatial model is fixed; see \S\ref{sec:LMC+MW_recovery}).  Parameters shown are the magnitude and phase of the $b_{\ell m}$ spherical harmonic coefficients up to $\blmax=2$ ($\almax=4$). Contours shown are 1- and $2\sigma$.}
    \label{fig:corners_lmc_mw_spatial}
\end{figure*}

Corner plots of the sampled posterior distributions for each of the analyses discussed are found on this and the following pages: Fig.~\ref{fig:corners_lmc_alone} for the LMC in isolation with LISA instrumental noise, and Fig.~\ref{fig:corners_lmc_mw_spectral} (Fig.~\ref{fig:corners_lmc_mw_spatial}) for the spectral (spatial) parameters of the analysis with the LMC + MW + LISA instrumental noise.


\bsp	
\label{lastpage}
\end{document}